\documentclass[a4paper]{jpconf}
\usepackage{amsfonts,amssymb,amsmath}  
\usepackage{array,dcolumn}             
\usepackage{longtable}                 
\usepackage{hyperref}                  
\usepackage[numbers,square,sort&compress]{natbib}
\usepackage{indentfirst}

\usepackage{commath}                   
\usepackage{xspace}                    
\usepackage{graphicx}  

\newcommand{\be}{\begin{equation}}
\newcommand{\ee}{\end{equation}}
\newcommand{\bea}{\begin{eqnarray}}
\newcommand{\eea}{\end{eqnarray}}                

\newcommand\alice{\textsc{alice}\xspace}

\newcommand\cms{\textsc{cms}\xspace}

\newcommand\fonll{\textsc{fonll}\xspace}

\newcommand\qgp{\textsc{qgp}\xspace}

\newcommand\pqcd{p\textsc{qcd}\xspace}

\newcommand\vusphydro{v-\textsc{usp}hydro\xspace}

\newcommand\pythia{\textsc{Pythia8}\xspace}

\newcommand\raa{\ensuremath{R_\text{AA}}\xspace}
\newcommand\pT{\ensuremath{p_T}\xspace}
\newcommand\vnn{\ensuremath{v_n}\xspace}
\newcommand\vn[1]{\ensuremath{v_{#1}}\xspace}

\newcommand\cum[1]{\ensuremath{\{#1\}}\xspace}

\newcommand\DD{D$^0$\xspace}

\begin{document}
\title{Heavy flavor \raa and \vnn in event-by-event viscous relativistic hydrodynamics}
\author{Caio~A.~G.~Prado$^1$, Jacquelyn Noronha-Hostler$^2$, Mauro R.~Cosentino$^3$, Marcelo G.~Munhoz$^1$, Jorge Noronha$^1$, Alexandre A.~P.~Suaide$^1$}
\address{$^1$ Instituto de F\'{i}sica, Universidade de S\~{a}o Paulo, C.P. 66318, 05315-970 S\~{a}o Paulo, SP, Brazil}
\address{$^2$ Department of Physics, University of Houston, Houston TX 77204, USA}
\address{$^3$ Centro de Ci\^{e}ncias Naturais e Humanas, Universidade Federal do ABC, Av. dos Estados 5001, Bairro Santa Terezina, 09210-508 Santo Andr\'{e}, SP, Brazil}
\ead{cagprado@if.usp.br}

\begin{abstract}
Recently it has been shown that a realistic description of the medium via event-by-event viscous hydrodynamics plays an important role in the long-standing $R_\text{AA}$ vs. $v_2$ puzzle at high $p_T$. In this proceedings we begin to extend this approach to the heavy flavor sector by investigating the effects of full event-by-event fluctuating hydrodynamic backgrounds on the nuclear suppression factor and $v_2\{2\}$ of heavy flavor mesons and non-photonic electrons at intermediate to high $p_T$. We also show results for $v_3\{2\}$ of $B^0$ and D$^0$ for PbPb collisions at $\sqrt{s}=2.76$ TeV.
\end{abstract}

\section{Introduction}

Heavy quarks, such as bottom and charm, are very useful probes of Quark-Gluon
Plasma (\qgp) dynamics. Because of their large mass, these quarks are
produced at the very early stages of the collision via hard processes and
their subsequent propagation through the hot and dense medium is sensitive to
the whole hydrodynamic evolution of the system. The energy loss experienced
by the heavy quarks during their path within the plasma has been studied using the nuclear modification factor (at mid-rapidity), defined as
\begin{equation}
  \raa(\pT,\varphi) = \frac{\frac{\dif N_\text{AA}}{\dif\pT\dif\varphi }}
                           {N_\text{coll}\frac{\dif N_\text{pp}}{\dif\pT }},
\end{equation}
where $\dif N_\text{AA}/\dif\pT=\frac{1}{2\pi}\int_0^{2\pi}d\varphi\frac{\dif N_\text{AA}}{\dif\pT\dif\varphi }$ is the spectrum of heavy quarks in AA collisions while $\dif N_\text{pp}/\dif\pT$ is the corresponding
proton-proton yield, $\varphi$ is the azimuthal angle in the plane transverse
to the beam direction, and $N_\text{coll}$ is the number of binary collisions
(computed within the Glauber model).

The dependence of $\raa(\pT,\varphi)$ on the azimuthal angle $\varphi$ can be
used to study the energy loss and its path length dependence in the plasma.
The degree of anisotropy in $\raa(\pT,\varphi)$ is determined via the
$\vnn^\text{heavy}$ coefficients of its Fourier expansion
\be
\frac{R_\text{AA}(p_T,\varphi)}{R_\text{AA}(p_T)} =  1  + 2\sum_{n=1}^\infty v_n^\text{heavy}(p_T) \cos\left[n\varphi - n\psi_n^\text{heavy}(p_T)  \right]
\ee
where $R_\text{AA}(p_T) =\frac{1}{2\pi} \int_0^{2\pi}d\varphi\,
R_\text{AA}(p_T,\varphi)$ is the azimuthal average and
\be
\label{definevnheavy}
v_n^\text{heavy}(p_T) = \frac{\frac{1}{2\pi}\int_0^{2\pi}d\varphi\,\cos\left[n\varphi-n\psi_n^\text{heavy}(p_T)\right]\,R_\text{AA}(p_T,\varphi)}{R_\text{AA}(p_T)}
\ee
and 
\be
\psi_n^\text{heavy}(p_T) = \frac{1}{n}\tan^{-1}\left(\frac{\int_0^{2\pi}d\varphi\,\sin\left(n\varphi\right)\,R_\text{AA}(p_T,\varphi)}{\int_0^{2\pi}d\varphi\,\cos\left(n\varphi\right)\,R_\text{AA}(p_T,\varphi)}\right).
\ee
As pointed out in~\cite{Noronha-Hostler2016,Betz:2016ayq}, quantities such as
$\vnn^\text{heavy}(p_T)$ do not actually correspond to what is measured. In
fact, just as it is done in the soft sector, the harmonic flow coefficients
either at high $p_T$ or in the heavy flavor sector are defined via
correlation functions between a soft particle (event plane) with a high $p_T$
particle or a heavy flavor candidate. This intrinsic correlation necessarily
requires~\cite{Noronha-Hostler2016,Betz:2016ayq} the use of event-by-event viscous
hydrodynamic simulations for the \qgp to correctly describe the underlying
flow, and its fluctuations, in the soft sector. Through such an approach one
can obtain a nonzero triangular flow at high $p_T$, as shown in~\cite{Noronha-Hostler2016,Betz:2016ayq}. 

In this proceedings we initiate the investigation of event-by-event viscous
hydrodynamic fluctuations on observables in the heavy flavor sector for PbPb
collisions at $\sqrt{s}=2.76$ TeV, as very briefly described below.

\section{Simulation}

We developed a new framework to describe the propagation of heavy quarks on
top of energy density and hydrodynamic flow profiles obtained via
event-by-event viscous hydrodynamics. The simulation follows a modular
paradigm, which allows for the separate study of different aspects of the
collision. We simulate the propagation of heavy quarks (bottom and charm) in
an expanding (boost invariant) medium described by the \vusphydro code~\cite{Noronha-Hostler2013b,Noronha-Hostler2014}
on an event-by-event basis for PbPb collisions at $\sqrt{s}=2.76$ TeV. Only
shear viscosity effects are taken into account and we assume $\eta/s=0.11$~\cite{Noronha-Hostler2016}.
\textsc{Mckln} initial conditions~\cite{Drescher2007} are used for the
hydrodynamic evolution. The system is evolved separately from the
heavy quark propagation, which are treated as probes, and thus we neglect
any effect of the probes on the medium, unlike \cite{Andrade2014a}. The hydrodynamic
evolution takes place until the complete freeze-out of the system, with the
freeze-out temperature set as $T_\text{FO}=140$ MeV, occurs. The hydrodynamic
simulation gives evolving profiles on the transverse plane for the energy
density $\varepsilon$ (and temperature $T$) and the components of the transverse
velocity $v_x$ and $v_y$ on an event-by-event basis.

Heavy quarks are sampled at the beginning of every hydrodynamic event (1000
events for each centrality class). The initial position of the quarks is
defined by the number of binary collisions in the event. We set a random
initial propagation direction $\varphi_\text{quark}$ and the initial momentum distribution
of the quarks is given by a \pqcd calculation (\fonll)~\cite{Cacciari1998}.
The heavy quarks lose energy to the evolving medium via the simple energy loss model~\cite{Betz2014}
\begin{equation}
  \dod{E}{x}= \alpha\Gamma_\text{flow}f(T,E,x),
\end{equation}
where
$\Gamma_\text{flow}=\gamma\left[1-v\cos(\varphi_\text{quark}-\varphi_\text{flow})\right]$
takes into account the local flow boost of the
medium~\cite{Noronha-Hostler2016}, $f(T,E,x)$ is a function that specifies
the energy loss model dependence with the medium temperature, heavy quark
energy $E$, and path length $x$. The coupling constant parameter $\alpha$
is found by comparison to data and here we use the \DD meson \raa spectrum for
central collisions to obtain it for the charm quark. After fixing the value
of $\alpha$ for charm, we take the electron contribution of both heavy
quarks and fit the parameter for the bottom quark using electron \raa data.
The heavy quarks propagate in the \qgp until they find a region where the
local temperature is smaller than a certain value, which we call the
jet-medium decoupling parameter, $T_d=120$ MeV, below which energy loss stops
and fragmentation is employed. We do not include coalescence effects in this
paper~\cite{Cao2016} and, thus, we will restrict ourselves to the high
$p_T$ region where these effects are minimal. The final electron spectra is
obtained from the meson decays, calculated using \pythia~\cite{SJOSTRAND2008}.
The calculation of the differential flow coefficients follows Ref.~\cite{Bilandzic2014} where here the soft and heavy flow harmonics are correlated, which is only possible in event-by-event calculations.

\section{Results}

\begin{figure} [h]
  \centering
  \includegraphics{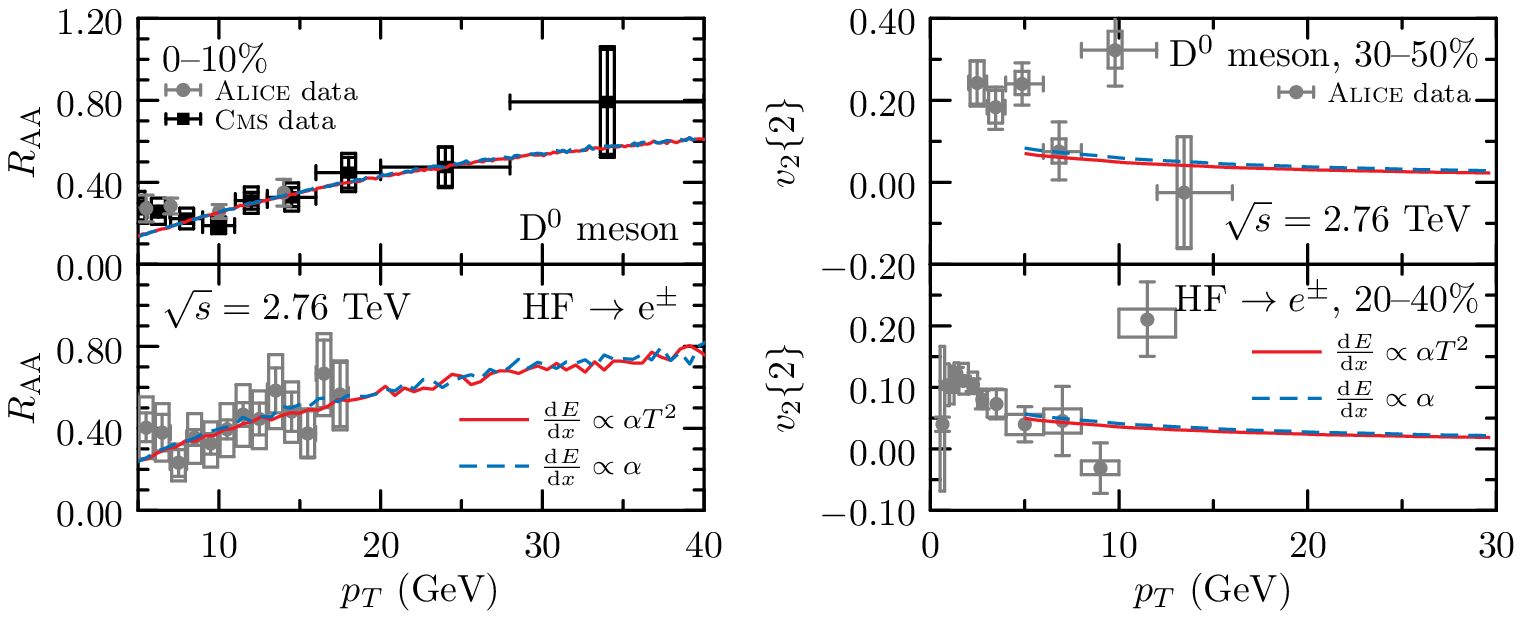}

  \begin{minipage}[t]{0.48\textwidth}
    \caption{\small\raa for \DD meson (top) and heavy flavor electron (bottom) for
    $\sqrt{s} = 2.76\text{ TeV}$ PbPb collisions and two
    different energy loss models. Data from \alice and
    \cms~\cite{ALICECollaboration2014c,CMS-PAS-HIN-15-005,ALICECollaboration2014d}
    are presented with the simulation results.}
    \label{fig:raa}
  \end{minipage}%
  \hfill
  \begin{minipage}[t]{0.48\textwidth}
    \caption{\small Differential $\vn2\cum2$ for \DD meson (top) and heavy flavor
    electron (bottom) for $\sqrt{s} = 2.76\text{ TeV}$ PbPb collisions and two
    different energy loss models. Data from
    the \alice experiment~\cite{ALICECollaboration2014c,ALICECollaboration2016}
    is also presented.}
    \label{fig:v2}
  \end{minipage}%
\end{figure}

In Fig.~\ref{fig:raa} our results for the nuclear modification factor are
compared to \DD meson and heavy flavor electron data for central PbPb
collisions at $\sqrt{s} =2.76\text{ TeV}$. The plot presents results for two
different energy loss models where $\dif E/\dif x \sim T^2$ and $\dif E/\dif
x = \text{constant}$. The 2-particle elliptic flow cumulant $\vn2\cum2$ for
non-central collisions is presented in Fig.~\ref{fig:v2} for \DD meson and
heavy flavor electrons.  At intermediate $p_T < 10$ GeV our results
are comparable to previous
calculations~\cite{Cao2016,Das2015} when coalescence effects are not
present. The two energy loss models that yielded the same nuclear
modification factor give different results for $\vn2\cum2$, which shows the
sensitivity of the azimuthal distribution with the energy loss mechanisms.

\begin{figure} [t]
  \centering
  \includegraphics{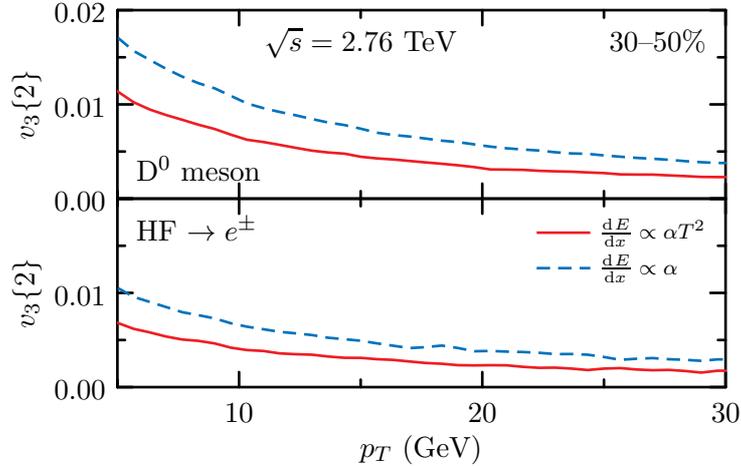}
  \caption{\small Differential $\vn3\cum2$ for \DD meson (top) and heavy flavor
  electron (bottom) for $\sqrt{s} = 2.76\text{ TeV}$ PbPb collisions and two
  different energy loss models.}
  \label{fig:v3}
\end{figure}



Fig.\ \ref{fig:v3} presents the first calculation of the 2-particle cumulant of triangular flow
$\vn3\cum2$ for \DD meson and heavy flavor electrons for non-central PbPb
collisions at $\sqrt{s} = 2.76\text{ TeV}$ for two energy loss models (we note that heavy flavor triangular flow, defined by the event plane method, has been computed in \cite{Nahrgang:2014vza}). We see that the distinction between the energy loss models is even more significant than what is found in the case of $\vn2\cum2$ in Fig.\ \ref{fig:v2} and, thus, $\vn3\cum2$ may be an even
better tool than $\vn2\cum2$ to learn about jet-medium interactions. 

\section{Conclusions}

We developed a new framework to study heavy probes in an event-by-event
hydrodynamically expanding viscous \qgp. Results for $\raa$ and $v_2\{2\}$ in the
heavy flavor sector from this approach were compared to available
experimental data. We also presented the first calculation of $\vn3\cum2$ for
heavy flavor, which revealed to be more sensitive to the choice of energy loss than $v_2\{2\}$. Future work includes the calculation of multiple-particle cumulants of harmonic flow and the inclusion of coalescence effects to also describe the intermediate
$p_T$ sector.

\ack
J.N.H. was supported by the National Science
Foundation under grant no. PHY-1513864. The authors thank \textsc{cnp}q and
\textsc{fapesp} for financial support.

\scriptsize
\bibliography{library}
\end{document}